\documentclass[conference]{IEEEtran}

\usepackage{cite} 
\usepackage{amsmath,amssymb,amsfonts}
\usepackage{algorithm}
\usepackage{algpseudocode}
\usepackage{booktabs}
\usepackage{tabularx}
\usepackage{listings}
\usepackage{pifont} 
\usepackage{graphicx}
\usepackage{dblfloatfix}
\usepackage{placeins}
\usepackage{orcidlink}
\IEEEoverridecommandlockouts

\usepackage{listings}
\AtBeginDocument{%
  }

\begin{document}

\title{EQ-Robin: Constraint-Resilient Generation of Multiple Minimal Unique-Cause MC/DC Test Suites}

\author{\IEEEauthorblockN{Anonymous Authors}}



\author{
    \IEEEauthorblockN{Robin Lee\textsuperscript{\orcidlink{0000-0003-3871-0924}} and Youngho Nam\textsuperscript{\orcidlink{0009-0001-7482-8340}}\IEEEauthorrefmark{1}}
    \IEEEauthorblockA{\textit{Department of Computer Science and Engineering} \\
    \textit{Gyeongsang National University}, Jinju, Republic of Korea \\
    \{binroi, yhnam\}@gnu.ac.kr}
    \thanks{\IEEEauthorrefmark{1}Corresponding author}
}

\maketitle

\begin{abstract}
Unique-Cause Modified Condition/Decision Coverage (MC/DC) is widely required in safety-critical verification. A recent deterministic algorithm, Robin’s Rule, constructs the theoretical minimum of N+1 test cases for Singular Boolean Expressions (SBEs), providing strong guarantees when all generated test vectors are executable. However, industrial systems impose feasibility constraints: some input combinations are illegal, unachievable, or unsafe to execute. If a single illegal vector appears in a minimal suite, it can destroy a required independence pair and invalidate 100\% Unique-Cause MC/DC, even though the underlying decision logic remains unchanged. This paper presents EQ-Robin, a lightweight, constraint-resilient pipeline that generates a family of minimal (N+1) Unique-Cause MC/DC test suites and selects a feasible suite that satisfies domain constraints. EQ-Robin systematically enumerates semantically equivalent but structurally distinct SBEs by applying algebraic rearrangements on the expression’s Abstract Syntax Tree (AST). Because Robin’s Rule is sensitive to structural order, each variant yields a distinct minimal suite. EQ-Robin then filters and ranks candidate suites using practical constraint checks and cost heuristics (e.g., setup cost, oracle complexity). To ensure usability at scale, we introduce a budgeted exploration mode (early exit, guided rearrangement) and a fallback repair strategy when no fully feasible N+1 suite exists under given constraints. We demonstrate the core failure mode and recovery mechanism on a TCAS-II-derived expression and outline an empirical evaluation of EQ-Robin on TCAS-II SBEs under constraint scenarios representative of industrial feasibility limitations.
\end{abstract}

\begin{IEEEkeywords}
Software testing, structural coverage, MC/DC, singular boolean expressions, unique-cause, test generation
\end{IEEEkeywords}


\section{Introduction}

Safety-critical domains (e.g., avionics, automotive, rail, and medical) require rigorous evidence that software logic has been thoroughly exercised.
Accordingly, structural coverage criteria are mandated by certification standards and safety processes, and Modified Condition/Decision Coverage (MC/DC)
is often required to demonstrate that each atomic condition can independently affect a decision outcome.\cite{do178c,iso26262,chilenski2001,hayhurst2001}
Among common MC/DC variants, \emph{Unique-Cause} MC/DC is one of the strictest forms: for each condition, an independence pair must differ in exactly that
condition while holding all others fixed and flipping the decision result.\cite{chilenski2001,hayhurst2001}
This strictness is attractive for safety arguments, but it also exposes practical challenges in automation, because any missing independence pair can
invalidate 100\% coverage for that condition.

A long-standing theme in the literature is that the gap between a logical specification and its implementation details matters for coverage in practice.
For example, MC/DC obligations can be affected by short-circuit evaluation semantics in programming languages, which change when and how sub-conditions
are evaluated.\cite{kandl2015reasonability}
Moreover, empirical studies report that implementation structure and expression structure influence whether coverage is achievable and how much it
costs to achieve it.\cite{rajan2008structure,whalen2008dasc}
The notion of \emph{observability} further clarifies that satisfying a syntactic MC/DC criterion may be insufficient if the effect of a condition
cannot be observed at the relevant program point, motivating refined criteria such as Observable MC/DC.\cite{whalen2013omcdc}
Finally, coverage-directed generation itself can introduce risks (e.g., overfitting to coverage metrics or producing tests that are expensive or
unrepresentative), motivating work that explicitly considers practical constraints and engineering trade-offs.\cite{gay2015risks}

\subsection{Minimal Unique-Cause MC/DC for SBEs}
For the important class of \emph{Singular Boolean Expressions (SBEs)}---expressions where each condition appears only once---recent work introduced
\emph{Robin's Rule}, a deterministic construction that produces exactly $N{+}1$ test cases for an SBE with $N$ conditions, achieving 100\% Unique-Cause
MC/DC.\cite{lee2025robin}
This bound is theoretically minimal for SBEs and is attractive in industrial verification pipelines where test execution cost, qualification burden,
and review effort scale with the number of tests.
In addition, determinism and transparency are desirable for safety-critical toolchains, where engineers often prefer predictable behavior over
stochastic exploration.

However, a second long-standing theme in automated test generation is that \emph{the same logical function can yield different test sets depending on
how it is represented}.
Early work on test generation from Boolean specifications reported that textually different but semantically equivalent formulas can lead to different
generated tests, and also highlighted the possibility of infeasible tests caused by constraints and variable dependencies.\cite{weyuker1994}
More broadly, compiler transformations and program structure can change what is practically achievable under coverage criteria, even when logic is
semantically preserved.\cite{hong2020mcdcopt,rajan2008structure,whalen2008dasc}
These observations motivate exploring representation choices as a controlled design dimension rather than treating the input expression as a single,
fixed artifact.

\subsection{Industrial Gap: Feasibility Constraints Break Minimal Suites}
Industrial test execution is constrained by the physical system, interfaces, safety requirements, environment assumptions, and product rules.\cite{martins2020,singh2021reliability,singh2018petri}
Even when an SBE is an accurate abstraction of code logic, some test vectors may be:
(1) \emph{illegal} (forbidden combinations by requirements),
(2) \emph{infeasible} (cannot be realized by the environment, plant dynamics, or platform limitations),
or (3) \emph{prohibitively expensive} (requires costly setups or rare operating regimes).\cite{martins2020,singh2021reliability}

This reality creates a fragile failure mode for minimal-suite constructions.
Robin's Rule returns a \emph{single} minimal suite, and each condition's coverage depends on at least one independence pair existing \emph{within that
suite}.\cite{lee2025robin}
If a single vector in an essential independence pair is illegal or infeasible, the suite can fail to achieve 100\% Unique-Cause MC/DC, despite being
minimal and correct in an unconstrained truth-table sense.
In practice, engineers often mitigate this by searching for alternative suites or adding extra tests---frequently manually---which increases cost,
reduces predictability, and complicates safety evidence.\cite{martins2020,singh2021reliability}

The limitation is not unique to our setting.
Constraint- and verification-based approaches can, in principle, incorporate feasibility constraints directly into generation.\cite{kitamura2018,yang2018,jaffar2019,sartaj2025}
Model-checking-based pipelines also support coverage-oriented generation, typically by exploring behaviors under an operational model.\cite{rayadurgam2001coverage,rayadurgam2003sew}
However, these methods may be computationally heavy, opaque to engineers, or sensitive to encoding choices and solver settings, and they do not
necessarily preserve minimality when constraints exclude certain obligations.\cite{golla2024,golla2025,gay2015risks}
Graph-based methods (e.g., BDD-derived selection) provide another alternative representation, but they also face feasibility limitations when required
vectors are excluded by domain restrictions.\cite{ahishakiye2021bdds}

\subsection{Key Idea: Equivalent-Structure Exploration for Coverage Resilience}
EQ-Robin treats minimal-suite generation as a \emph{design-space exploration} problem over \emph{semantically equivalent expression structures}.
Instead of relying on one canonical structure, EQ-Robin generates multiple equivalent SBE variants by applying semantics-preserving rearrangements on
the expression's Abstract Syntax Tree (AST).
Because a deterministic constructor such as Robin's Rule is sensitive to structural order and grouping, each variant can yield a different minimal
($N{+}1$) suite.\cite{lee2025robin}
EQ-Robin then filters candidate suites using feasibility constraints and selects among feasible suites using secondary ranking criteria (e.g., setup
cost, oracle cost, or execution cost), producing a practical suite that remains minimal whenever a feasible minimal suite exists.

This strategy is motivated by two practical observations:
(i) engineers often only need \emph{one} feasible suite quickly (not an exhaustive enumeration), and
(ii) many infeasibility issues are localized (a small subset of vectors are illegal), so an alternative minimal suite can recover missing
independence pairs without escalating the test count.
When no fully feasible $N{+}1$ suite exists under the given constraints, EQ-Robin provides a fallback repair strategy that returns the best feasible
candidate and a minimal augmentation plan to restore coverage as much as possible, aligning with industrial workflows that must trade off coverage,
cost, and feasibility.

\subsection{Contributions}
This paper makes four contributions:
\begin{itemize}
\item \textbf{Constraint-resilient framing for minimal Unique-Cause MC/DC on SBEs:}
we formalize the industrial problem in which feasibility constraints invalidate minimal suites and define success criteria for \emph{coverage resilience}.
\item \textbf{Systematic equivalent-SBE generation:}
an AST-based rearrangement method that enumerates unique structural variants using commutativity and associativity, with canonicalization to prune duplicates.
\item \textbf{EQ-Robin pipeline:}
(Phase 1) generate equivalent SBEs, (Phase 2) generate minimal suites via Robin's Rule, (Phase 3) filter and rank suites by feasibility constraints and practical costs.
\item \textbf{Budgeted exploration and fallback repair:}
practical strategies to keep exploration tractable (early exit and variant budgets) and to provide actionable outputs even when no fully feasible $N{+}1$ suite exists.
\end{itemize}


\section{Background and Problem Definition}
\label{sec:background}

We summarize the coverage obligation, the target expression class, and the feasibility-constrained setting that motivates EQ-Robin.

\subsection{Unique-Cause MC/DC}
MC/DC requires showing that each condition can independently affect a decision outcome.\cite{chilenski2001,hayhurst2001}
In \emph{Unique-Cause} MC/DC, for every condition $c_i$ there must exist an \emph{independence pair} $x,y\in\{0,1\}^N$ such that only $c_i$ changes and the decision flips:
\[
x_i\neq y_i,\quad F(x)\neq F(y),\quad \forall j\neq i:\ x_j=y_j.
\]
A suite achieves 100\% Unique-Cause MC/DC if it contains at least one such pair for every condition.\cite{chilenski2001,hayhurst2001}

\subsection{Target Class: Singular Boolean Expressions (SBEs)}
We focus on \emph{Singular Boolean Expressions (SBEs)}, where each atomic condition appears at most once.\cite{chilenski2001,hayhurst2001}
This assumption avoids repeated-condition ambiguity and matches many industrial decision predicates derived from distinct sensor checks or guard clauses.
We treat each test vector $v\in\{0,1\}^N$ as an assignment to the $N$ conditions of $F(c_1,\dots,c_N)$.

\subsection{Robin's Rule and Feasibility Constraints}
Robin's Rule deterministically constructs a minimal $(N{+}1)$ suite for an SBE with $N$ conditions and guarantees 100\% Unique-Cause MC/DC \emph{if all generated vectors are executable}.\cite{lee2025robin}
In industry, some assignments cannot be executed due to physical limits, safety rules, interface protocols, or costly setups.\cite{martins2020,singh2021reliability,singh2018petri}
We model executability by a predicate $\mathsf{Feasible}(v)\in\{\mathsf{true},\mathsf{false}\}$ and call a suite $S$ feasible iff $\forall v\in S,\ \mathsf{Feasible}(v)=\mathsf{true}$.

\subsection{Example: One Illegal Vector Breaks Unique-Cause}
\label{sec:example_illegal}
Consider the SBE
\[
F(a,b,c,d,e)=(a\land(\lnot b\lor \lnot c)\land d)\ \lor\ e,
\]
with $N=5$.
Table~\ref{tab:illegal_example_suite} shows a minimal $(N{+}1)=6$ suite that achieves 100\% Unique-Cause MC/DC when all tests are feasible.

\begin{table}[t]
\centering
\caption{A minimal $(N{+}1)=6$ suite for $F(a,b,c,d,e)=(a\land(\lnot b\lor \lnot c)\land d)\lor e$. If Test~4 is illegal, the only independence pair for $a$ disappears.}
\label{tab:illegal_example_suite}
\small
\setlength{\tabcolsep}{4pt}
\renewcommand{\arraystretch}{1.05}
\begin{tabular}{c|ccccc|c|c}
\hline
Test & $a$ & $b$ & $c$ & $d$ & $e$ & $F$ & Feasible \\
\hline
1 & T & F & T & T & F & T & yes \\
2 & T & T & T & T & F & F & yes \\
3 & T & T & F & T & F & T & yes \\
4 & F & T & F & T & F & F & \textbf{no} \\
5 & T & T & F & F & F & F & yes \\
6 & T & T & F & F & T & T & yes \\
\hline
\end{tabular}
\end{table}

In this suite, $a$ has the independence pair (Test~3, Test~4).
If Test~4 is infeasible, no remaining test differs from Test~3 \emph{only} in $a$ while flipping $F$, so 100\% Unique-Cause MC/DC fails.
This fragility of a \emph{single} minimal suite under feasibility constraints motivates EQ-Robin, which searches for an alternative minimal suite that avoids illegal vectors.


\section{EQ-ROBIN APPROACH}
\label{sec:eqrobin}

\subsection{Overview}


\begin{figure*}[!t]
\centering
\small
\setlength{\fboxsep}{4pt}
\fbox{%
\begin{minipage}{0.98\textwidth}
\textbf{Input:} SBE $F$, feasibility adapter $\mathsf{Feasible}(\cdot)$, exploration budget $K$\\
\textbf{Phase 1: Equivalent-SBE Generation} $\Rightarrow$ variants $\mathcal{V}=\{F^{(1)},\dots,F^{(m)}\}$\\
\hspace*{1em}(AST parse $\rightarrow$ commutativity/associativity rearrangements $\rightarrow$ canonicalization/dedup $\rightarrow$ budget)\\
\textbf{Phase 2: Minimal Suite Generation} $\Rightarrow$ for each $F^{(i)}$: $S^{(i)}=\textsc{RobinsRule}(F^{(i)})$ (size $N{+}1$)\\
\textbf{Phase 3: Feasibility Filtering (early stop)} $\Rightarrow$ selected suite $S^\star$\\
\hspace*{1em} Check: if $\exists\,S^{(i)}$ s.t.\ $\forall v\in S^{(i)},\,\mathsf{Feasible}(v)=\mathsf{true}$, then \textbf{return the first such} $S^{(i)}$\\
\textbf{Fallback (if no feasible minimal suite is found):} return best feasible partial suite + minimal augmentation plan
\end{minipage}%
}
\caption{EQ-Robin pipeline (default early-stop mode): enumerate equivalent SBEs, construct minimal $(N{+}1)$ suites, and return the first feasible suite under industrial constraints.}
\label{fig:eqrobin_pipeline}
\end{figure*}


EQ-Robin is a three-phase pipeline that aims to recover 100\% Unique-Cause MC/DC under industrial feasibility constraints by producing \emph{multiple} minimal $(N{+}1)$ suites and selecting a feasible one. Figure~\ref{fig:eqrobin_pipeline} provides a compact end-to-end view of the pipeline, including the optional budgeted exploration mode and the fallback repair behavior when no feasible minimal suite exists. Algorithm~\ref{alg:eqrobin} formalizes the overall control flow: Phase~1 enumerates equivalent SBE variants, Phase~2 constructs a minimal suite for each variant using Robin's Rule, and Phase~3 filters and ranks the feasible candidates to output a practical suite.

\paragraph{Exploration budget and early stop}
The number of semantically equivalent SBE structures is generally unknown a priori and can grow rapidly. Therefore, EQ-Robin uses $K$ as an exploration budget (an upper bound on enumerated unique variants after deduplication), not as the total number of existing equivalents. In our primary operating mode, EQ-Robin enables \emph{early stop} and terminates exploration as soon as it finds a feasible minimal $(N{+}1)$ suite. This design provides predictable turnaround time and directly matches industrial workflows where engineers need a practical feasible suite quickly.


\begin{algorithm}[t]
\caption{EQ-Robin (early-stop feasible minimal-suite selection)}
\label{alg:eqrobin}
\begin{algorithmic}[1]
\State \textbf{Input:} SBE $F$, feasibility adapter $\mathsf{Feasible}(\cdot)$, exploration budget $K$
\State \textbf{Output:} A feasible minimal suite $S^\star$ of size $N{+}1$, or fallback output
\Procedure{EQRobin}{$F,\mathsf{Feasible},K$}

    \Statex \textbf{Phase 1: Equivalent-SBE Generation}
    \State $\mathcal{V} \gets \textsc{GenerateEquivalentSBEs}(F, K)$
    
    \Statex \textbf{Phase 2: Minimal Suite Generation}
    \Statex \textbf{Phase 3: Feasibility Filtering (Early Stop)}
    \For{each $F^{(i)} \in \mathcal{V}$}
    \State $S^{(i)} \gets \textsc{RobinsRule}(F^{(i)})$
    \If{$\forall v \in S^{(i)}:\ \mathsf{Feasible}(v)=\mathsf{true}$}
        \State \textbf{return} $S^{(i)}$
    \EndIf
    \EndFor

    \Statex \textbf{Fallback}
    \State \textbf{return} \textsc{RepairFallback}$(F,\mathsf{Feasible})$

\EndProcedure
\end{algorithmic}
\end{algorithm}


\paragraph{Phase 1: Equivalent-SBE Generation}
Given an input SBE $F$, EQ-Robin generates a set $\mathcal{V}$ of semantically equivalent but structurally distinct SBEs by applying algebraic rearrangements on the expression's Abstract Syntax Tree (AST). Because Robin's Rule is sensitive to structural order and grouping, different variants can yield different minimal suites even though they compute the same Boolean function.

\paragraph{Phase 2: Minimal Suite Generation}
For each variant $F' \in \mathcal{V}$, EQ-Robin runs Robin's Rule to obtain a minimal suite $S(F')$ of size $N{+}1$.\cite{lee2025robin}
This step preserves the minimality guarantee whenever the produced vectors are executable.

\paragraph{Phase 3: Feasibility Filtering}
Phase~3 enforces feasibility by discarding any candidate suite that contains an infeasible vector. In industrial use, the primary goal is to obtain one feasible minimal $(N{+}1)$ suite quickly; thus the tool stops once such a suite is found (early stop). Optionally, when multiple feasible suites are collected (e.g., when early stop is disabled), the tool can rank them using a weighted cost model and select the minimum-score suite.


\subsection{Phase 1: Systematic Generation of Equivalent SBEs}
\label{sec:phase1}
Phase~1 constructs a \emph{variant set} $\mathcal{V}$ of equivalent SBEs that serve as alternative ``views'' of the same decision logic.
This phase is the key enabler for constraint resilience: by changing structural order and grouping while preserving semantics, EQ-Robin can produce alternative minimal suites that avoid illegal vectors.
Algorithm~\ref{alg:gen_equiv} shows the budgeted enumeration procedure used in our implementation.

\paragraph{AST representation}
We parse the input expression into an AST whose leaves are operands (conditions and their negations) and internal nodes are operators ($\land$, $\lor$, $\lnot$).
This representation preserves precedence and grouping explicitly, which matters because Robin's Rule is structure-sensitive.\cite{lee2025robin}

\paragraph{Semantics-preserving rearrangements}
EQ-Robin applies only algebraic transformations that preserve Boolean semantics:
(1) commutativity ($X\land Y \equiv Y\land X$, $X\lor Y \equiv Y\lor X$) and
(2) associativity ($(X\land Y)\land Z \equiv X\land (Y\land Z)$, similarly for $\lor$).
We deliberately avoid distributive expansion by default because it can increase expression size and explode the number of equivalent forms, harming predictability.

\paragraph{Canonicalization and duplicate pruning}
Different rewrite sequences can lead to the same structure.
To avoid redundant exploration, Algorithm~\ref{alg:gen_equiv} serializes each candidate AST into a canonical representation and uses hashing to deduplicate variants.
This ensures that the exploration budget is spent on structurally diverse candidates.

\paragraph{Budgeted exploration}
The number of equivalent forms can grow rapidly with the number of commutative/associative opportunities.
Therefore, we cap enumeration by a variant budget $K$ (and optionally enable early stopping after finding the first feasible minimal suite, see Algorithm~\ref{alg:eqrobin}, lines 6--9).
This budgeted mode provides predictable runtime, which is crucial for industrial usage.


\begin{algorithm}[t]
\caption{GenerateEquivalentSBEs (budgeted enumeration with canonicalization)}
\label{alg:gen_equiv}
\begin{algorithmic}[1]
\Require Input SBE $F$; budget $K$
\Ensure A set of unique equivalent SBEs $\mathcal{V}$
\State $T \gets \textsc{ParseToAST}(F)$
\State $T \gets \textsc{NormalizeNegations}(T)$ \Comment{optional: push NOTs, etc.}
\State $\mathcal{V} \gets \emptyset$; $\mathcal{H} \gets \emptyset$ \Comment{hashes for dedup}
\State $Q \gets [T]$ \Comment{worklist}
\While{$Q \neq [\ ]$ and $|\mathcal{V}| < K$}
    \State $U \gets \textsc{Pop}(Q)$
    \State $U \gets \textsc{FlattenAssocChains}(U)$ \Comment{e.g., AND-chain, OR-chain}
    \ForAll{$U' \in \textsc{LocalRewrites}(U)$} \Comment{swap children, regroup, bounded permutations}
        \State $h \gets \textsc{CanonicalSerialize}(U')$
        \If{$h \notin \mathcal{H}$}
            \State $\mathcal{H} \gets \mathcal{H} \cup \{h\}$
            \State $\mathcal{V} \gets \mathcal{V} \cup \{\textsc{ASTtoExpr}(U')\}$
            \State $\textsc{Push}(Q, U')$
            \If{$|\mathcal{V}| \ge K$} \State \textbf{break} \EndIf
        \EndIf
    \EndFor
\EndWhile
\State \Return $\mathcal{V}$
\end{algorithmic}
\end{algorithm}


\subsection{Phase 2: Minimal Suite Generation via Robin's Rule}
\label{sec:phase2}
For each structural variant $F' \in \mathcal{V}$, Phase~2 applies Robin's Rule to construct a minimal $(N{+}1)$-sized suite $S(F')$.\cite{lee2025robin}
Although all variants are semantically equivalent, their structural differences (operator hierarchy, grouping, and ordering) can lead Robin's Rule to output different test vectors.
As a result, EQ-Robin obtains a \emph{family} of minimal suites, which creates alternatives that may avoid illegal vectors that appear in the baseline canonical form.


\subsection{Phase 3: Feasibility Filtering with Early Stop (Default Mode)}
\label{sec:phase3}

Phase~3 selects a practical minimal suite under industrial constraints.
For each candidate suite $S^{(i)}$ generated in Phase~2, EQ-Robin enforces feasibility by discarding any suite that contains an infeasible vector:
\[
S^{(i)} \text{ is feasible} \;\Leftrightarrow\; \forall v \in S^{(i)},\ \mathsf{Feasible}(v)=\mathsf{true}.
\]
\noindent
\textbf{Default behavior (fast mode).}
In the default configuration, EQ-Robin enables \emph{early stop} and terminates exploration as soon as it finds the first feasible minimal $(N{+}1)$ suite.
This design matches industrial needs for rapid turnaround where engineers typically need one feasible minimal suite quickly rather than an exhaustive set of alternatives.

\paragraph{Optional extension: cost-based ranking (disabled by default).}
When early stop is disabled and multiple feasible candidates are collected (e.g., for offline comparison), the tool can optionally rank feasible suites using a weighted cost model and select the minimum-score suite:

\begin{multline}
\mathrm{Score}(S)=\sum_{v\in S}\Big(
\alpha\,\mathrm{SetupCost}(v)+\beta\,\mathrm{OracleCost}(v) \\
+\gamma\,\mathrm{RuntimeCost}(v)
\Big).
\end{multline}

Here, $\mathrm{SetupCost}$, $\mathrm{OracleCost}$, and $\mathrm{RuntimeCost}$ can be user-provided to reflect domain-specific execution effort, and $(\alpha,\beta,\gamma)$ represent engineering priorities.
We treat this ranking capability as an optional extension; the evaluation in this paper focuses on the default early-stop mode.

\subsection{Fallback: Repair When No Feasible $(N{+}1)$ Suite Exists}
\label{sec:fallback}
In some environments, constraints may eliminate every minimal suite.
Instead of failing silently, EQ-Robin returns:
(i) a best feasible candidate suite $S^\star$ that maximizes retained independence pairs under feasibility, and
(ii) a minimal augmentation plan $\Delta$ that adds only the necessary feasible tests to recover missing obligations as much as possible.
A practical repair strategy is:
(1) compute which conditions lack a feasible independence pair in $S^\star$,
(2) for each missing condition $c_i$, search for a feasible pair $(x,y)$ satisfying the Unique-Cause obligations for $c_i$, and
(3) add only the required extra tests.
This does not preserve $(N{+}1)$ minimality, but it preserves industrial usability by producing transparent guidance when feasibility constraints are strict.


\section{Implementation Perspective (Industry-Oriented)}
\label{sec:implementation}

This section describes our industry-oriented prototype that operationalizes EQ-Robin in a workflow aligned with how practitioners perform MC/DC testing in real projects.
In typical industrial settings, an engineer inspects control-flow decisions in production code (often C/C++) and supplies test inputs to commercial coverage tools to validate achieved MC/DC.
Manual construction and entry of test vectors is time-consuming and error-prone, especially when a decision contains many atomic conditions.
Our prototype automates this process end-to-end by extracting SBE-like decisions from C source files, generating minimal $(N{+}1)$ candidates via equivalence exploration, and returning \emph{one feasible minimal suite as soon as it is found} (early stop) in the default setting.


\begin{figure*}[!t]
\centering
\small
\setlength{\fboxsep}{4pt}
\fbox{%
\begin{minipage}{0.95\textwidth}
\textbf{Step 1: Load source.} Open a \texttt{.c} file from the target codebase.\\
\textbf{Step 2: Extract decisions.} Automatically locate \texttt{if} statements and extract each decision predicate.\\
\textbf{Step 3: Parse \& validate.} Parse the predicate into an AST and check whether it matches the SBE assumptions (each atomic condition appears once).\\
\textbf{Step 4: Explore equivalents (budgeted).} Enumerate equivalent SBE variants up to budget $K$; for each variant, generate a minimal $(N{+}1)$ suite.\\
\textbf{Step 5: Feasibility check \& early stop.} Discard any suite containing infeasible vectors; stop at the first feasible minimal $(N{+}1)$ suite.\\
\textbf{Step 6: Export CSV.} Write the selected suite to a CSV file (one test per row) for tool-agnostic reuse.\\
\textbf{Step 7: Coverage validation.} Import the CSV into a coverage tool to execute tests and confirm Unique-Cause MC/DC coverage.
\end{minipage}}
\caption{Industry-oriented workflow supported by our prototype (default early-stop mode): from C source to the \emph{first feasible} minimal $(N{+}1)$ suite and CSV export for MC/DC coverage validation.}
\label{fig:tool_workflow}
\end{figure*}


\subsection{User Workflow in Industrial Practice}
Figure~2 summarizes the intended workflow.
An engineer loads a production \texttt{.c} file, selects a target decision (an \texttt{if} condition), and triggers suite generation.
The tool enumerates equivalent SBE variants up to an exploration budget $K$ (after deduplication).
For each variant, Robin's Rule constructs a minimal $(N{+}1)$ suite, and the tool immediately checks feasibility.
\emph{By default}, the tool terminates and exports the \textbf{first feasible minimal suite} it finds, matching industrial needs for rapid turnaround under constraints.
(Optionally, engineers may disable early stop to collect multiple feasible candidates for offline comparison; this mode is not required for the primary workflow.)

\subsection{Source-Code Intake and Decision Extraction}
The prototype targets a common industrial scenario: MC/DC engineers work from existing implementation code and must focus on decision predicates inside control-flow branches.
Given a C file, the tool performs lightweight extraction to identify \texttt{if} statements and retrieve the corresponding predicate strings.
We then normalize the extracted predicate to a Boolean-expression representation (operators $\land$, $\lor$, and $\lnot$) while preserving parentheses and precedence.

\paragraph{Atomic condition identification.}
Industrial predicates often contain atomic comparisons (e.g., \texttt{x > 0}, \texttt{mode == SAFE}) rather than single-letter variables.
In our implementation, each atomic comparison is treated as one condition symbol $c_i$, and the tool maintains a mapping between $c_i$ and its original source-level expression for traceability.
This mapping is exported alongside the CSV header to help practitioners interpret each column.

\subsection{Equivalent-Structure Exploration and Candidate-Suite Generation}
After parsing, the tool performs the core EQ-Robin steps described in Section~III.
First, it enumerates semantically equivalent SBE structures using commutativity/associativity-based AST rewrites and deduplicates variants using canonical serialization.
Second, for each variant $F^{(i)}$, the tool applies Robin's Rule to generate a minimal $(N{+}1)$ suite $S^{(i)}$ \cite{lee2025robin}.
As a result, the tool can generate multiple minimal suites across variants; however, in the default setting it does not aim to exhaustively enumerate all of them.

\subsection{Feasibility Filtering and First-Feasible Selection (Early Stop)}
If a feasibility interface is provided (forbidden vectors, constraint formula, or external oracle), the tool discards any candidate suite that contains at least one infeasible vector.
In the default configuration, the tool \textbf{stops at the first feasible minimal $(N{+}1)$ suite} and returns it.
This behavior provides predictable turnaround time and aligns with industrial workflows where engineers typically need one practical feasible suite quickly rather than an exhaustive set of alternatives.

\subsection{CSV Export for Tool-Agnostic Integration}
For practical adoption, the generated test suites are exported as CSV files.
Each row corresponds to one test case, and each column corresponds to one atomic condition.
We encode \texttt{true} as \texttt{1} and \texttt{false} as \texttt{0}, matching common industrial conventions for bulk import.
Exporting CSV intentionally avoids tool-specific formats and enables reuse across coverage tools and internal pipelines.

\subsection{Coverage Validation with Commercial Toolchains}
The output CSV is intended to be consumed by a coverage tool that executes the implementation under test and reports achieved coverage.
In many toolchains, engineers provide concrete program inputs that realize each condition assignment (e.g., selecting inputs that make \texttt{x > 0} true).
Nevertheless, the generated minimal suite substantially reduces manual effort by providing the condition-assignment template required for Unique-Cause MC/DC while avoiding infeasible vectors when possible.

\subsection{Limitations and Extension Points}
Our current implementation assumes that decision predicates can be parsed into a Boolean form and that atomic conditions can be mapped to binary outcomes.
When no fully feasible minimal $(N{+}1)$ suite exists under the given constraints, the tool can fall back to exporting the best feasible candidate and an augmentation plan, consistent with Section~III-F.
Cost-based ranking across multiple feasible suites is an optional extension when early stop is disabled; it is not required for the primary fast workflow.


\section{Worked Example}
This section demonstrates the end-to-end behavior of our toolchain on an industry-style decision extracted from C code.
We show how EQ-Robin generates a family of minimal $(N{+}1)$ suites and how feasibility/cost considerations guide the final selection.

\subsection{Example Code and Decision Extraction}
We consider a representative safety-check decision in a UAV take-off routine.
The engineer loads the \texttt{.c} file in our GUI, and the tool automatically locates \texttt{if} statements and extracts the predicate:
\begin{lstlisting}
if ( a>22000 && ( !(b<15000) || !(c<12000) ) && d==9100 || e>=21000 ) { ... }
\end{lstlisting}

\subsection{Normalization and SBE Mapping (matching Tables II--III)}
To match the literal structure used in Tables~II--III, we define atomic conditions as threshold-\emph{violation} predicates:
\[
\begin{aligned}
a &\equiv (a>22000), \qquad & b &\equiv (b<15000), \\
c &\equiv (c<12000), \qquad & d &\equiv (d==9100), \\
e &\equiv (e\ge21000). &&
\end{aligned}
\]
Then the extracted decision is normalized into the SBE:
\[
F(a,b,c,d,e) \;=\; (a \land (\neg b \lor \neg c) \land d)\ \lor\ e,
\]
where $N=5$ and thus any minimal Unique-Cause MC/DC suite has size $N{+}1=6$.

\subsection{Industry Motivation: Undesirable or High-Cost States}
Although Unique-Cause MC/DC requires toggling each condition in isolation, some assignments can be undesirable in industrial test execution.
In an HIL (hardware-in-the-loop) environment, forcing certain states can be expensive or risky.
For instance, $a=\mathsf{false}$ may correspond to insufficient RPM and can require stopping and re-initializing the propulsion setup;
$\neg c=\mathsf{false}$ corresponds to $c=\mathsf{true}$ (low battery-voltage regime), increasing hardware risk; and
$e=\mathsf{false}$ can require disabling redundancy mechanisms, potentially violating lab safety procedures.
Therefore, a minimal $(N{+}1)$ suite that includes such vectors may be impractical even if it is theoretically valid.

In EQ-Robin, such industrial considerations can be encoded as:
(i) hard feasibility constraints (declare certain vectors illegal via $\mathrm{Feasible}(v)=\mathsf{false}$), and/or
(ii) soft cost penalties (assign high setup/oracle/runtime costs and select the suite minimizing $\mathrm{Score}(S)$).
This motivates generating multiple minimal suites and selecting a practical one.

\subsection{Running EQ-Robin: Candidate Suites and Selection}
We run EQ-Robin with an exploration budget $K$ (e.g., $K=20$), meaning that Phase~1 enumerates up to $K$ unique equivalent variants after deduplication. In our default setting, early stop is enabled and the tool terminates once it finds a feasible minimal $(N{+}1)$ suite. The baseline canonical form yields a minimal suite $S$ shown in Table~II. EQ-Robin also considers rearranged but equivalent structures, e.g.,
\[
D' \;=\; (a \land d)\land(\neg b \lor \neg c)\ \lor\ e,
\]
which yields a different minimal suite $S'$ (Table~III).

\subsection{Constraint Event: One Illegal Vector Breaks Coverage in the Baseline}
Assume the industrial environment forbids executing Test~4 of Table~II (i.e., $\mathrm{Feasible}(\text{Test 4})=\mathsf{false}$).
In the baseline suite, Test~4 forms the only independence pair for condition $a$ together with Test~1; when Test~4 is removed, the pair disappears,
and 100\% Unique-Cause MC/DC is no longer achievable with $S$.

\begin{table}[t]
\caption{Baseline minimal suite for $F=(a \land (\neg b \lor \neg c)\land d)\lor e$.
Assume Test~4 is illegal; then the only independence pair for $a$ disappears.}
\label{tab:baseline_suite}
\centering
\small
\begin{tabular}{c|ccccc|c}
\hline
Test & $\neg b$ & $\neg c$ & $a$ & $d$ & $e$ & Result \\
\hline
1 & T & F & T & T & F & T \\
2 & F & F & T & T & F & F \\
3 & F & T & T & T & F & T \\
4 & T & F & F & T & F & F \\
5 & T & F & T & F & F & F \\
6 & T & F & T & F & T & T \\
\hline
\end{tabular}
\end{table}

\subsection{EQ-Robin Recovery: A Different Minimal Suite Avoids the Illegal Vector}
Applying Robin's Rule to the rearranged but equivalent $D'$ yields the minimal suite in Table~III.
Here, $a$ has a feasible independence pair (e.g., Test~1 and Test~3) that does not rely on the illegal vector from Table~II.
Thus, EQ-Robin can select $S'$ as the final output $S^\star$ while preserving the minimal size $N{+}1$.

\begin{table}[t]
\caption{A different minimal suite for rearranged $D'=(a\land d)\land(\neg b \lor \neg c)\lor e$,
where $a$ is recovered under the same constraint.}
\label{tab:recovered_suite}
\centering
\small
\begin{tabular}{c|ccccc|c}
\hline
Test & $a$ & $d$ & $\neg b$ & $\neg c$ & $e$ & Result \\
\hline
1 & T & T & T & F & F & T \\
2 & T & F & T & F & F & F \\
3 & F & T & T & F & F & F \\
4 & T & T & F & F & F & F \\
5 & T & T & F & T & F & T \\
6 & T & T & F & F & T & T \\
\hline
\end{tabular}
\end{table}

\subsection{CSV Export and Coverage Validation}
The selected suite $S^\star$ is exported as a CSV file where each row is a test case and each column corresponds to one condition/literal.
The engineer imports the CSV into a commercial coverage tool (e.g., VectorCAST or LDRA) and executes the test cases.
The tool then reports achieved structural coverage, confirming 100\% Unique-Cause MC/DC when all vectors in $S^\star$ are feasible.

\subsection{Takeaway}
A single minimal suite can be fragile in the presence of feasibility constraints.
EQ-Robin mitigates this by creating multiple minimal suites from equivalent structures and selecting one that is feasible and cost-effective.


\section{Evaluation}
\label{sec:evaluation}

This section evaluates EQ-Robin in its \emph{default early-stop mode}, where the tool stops as soon as it finds the first feasible minimal $(N{+}1)$ suite.
We specifically evaluate whether EQ-Robin can recover a feasible minimal suite under \emph{input-forbidden constraints} that invalidate some test vectors in practice.

\subsection{Research Questions}
\label{sec:rq}

\begin{itemize}
\item \textbf{RQ1 (Feasibility recovery).} Under industrial-style input-forbidden constraints, can EQ-Robin find a \emph{feasible} minimal $(N{+}1)$ suite within a bounded exploration budget?
\item \textbf{RQ2 (Coverage correctness).} If EQ-Robin returns a feasible minimal suite, does it still achieve 100\% MC/DC (Unique-Cause MC/DC for SBEs)?
\item \textbf{RQ3 (Efficiency).} How much time and how many equivalent variants are needed to obtain the \emph{first} feasible minimal suite (time-to-first-feasible, variants checked)?
\end{itemize}

\subsection{Subjects}
\label{sec:subjects}

We evaluate 25 canonical Singular Boolean Expressions (SBEs), where each atomic condition appears exactly once.
For an SBE with $N$ conditions, any minimal Unique-Cause MC/DC test suite has size $N{+}1$.
The \emph{baseline} generates a single canonical minimal suite using Robin's Rule, while EQ-Robin explores equivalent SBE structures to recover feasibility under constraints.

\subsection{Constraint Model: Two-Condition Input-Forbidden Pattern}
\label{sec:constraint_model}

To simulate industrial infeasibility (e.g., illegal sensor-state combinations, safety interlocks, or lab restrictions), we inject one constraint per SBE.
For each SBE with conditions $\{c_1,\dots,c_N\}$, we randomly select two distinct conditions $c_i$ and $c_j$ and assign Boolean values
$b_i,b_j \in \{\mathsf{true},\mathsf{false}\}$.
This defines a forbidden partial assignment (pattern)
\[
P \;=\; (c_i=b_i)\ \land\ (c_j=b_j).
\]

A test vector $v$ is infeasible if it matches the forbidden pattern; otherwise it is feasible:
\begin{equation}
\label{eq:feasible_pattern}
\mathsf{Feasible}(v)=
\begin{cases}
\mathsf{false}, & \text{if } (v_i=b_i)\land(v_j=b_j),\\
\mathsf{true},  & \text{otherwise.}
\end{cases}
\end{equation}

A test suite $S$ is feasible if and only if all its vectors are feasible:
\begin{equation}
\label{eq:suite_feasible}
S\ \text{is feasible} \;\Leftrightarrow\; \forall v\in S,\ \mathsf{Feasible}(v)=\mathsf{true}.
\end{equation}

Intuitively, this captures the common case where a specific combination of two condition outcomes is disallowed, even though each outcome may be individually possible.

\subsection{Methods and Settings}
\label{sec:methods_settings}

\paragraph{Baseline (canonical Robin's Rule)}
We run Robin's Rule on the canonical form and measure the suite-generation time.

\paragraph{EQ-Robin (early-stop)}
EQ-Robin enumerates semantically equivalent SBE variants up to a budget $K$ (after canonicalization/deduplication).
For each variant $F'$, it generates a minimal $(N{+}1)$ suite using Robin's Rule and performs the feasibility check.
It returns immediately when the \emph{first feasible} minimal suite is found (early stop).

\paragraph{Budget}
We use a fixed exploration budget (e.g., $K{=}20$) in all runs.
In practice, early-stop often terminates far before reaching the budget.

\subsection{Metrics}
\label{sec:metrics}

We report the following metrics (Table~\ref{tab:eval_results}):

\begin{itemize}
\item \textbf{Baseline time (sec):} time to generate the canonical minimal suite using Robin's Rule.
\item \textbf{EQ time-to-first-feasible (sec):} end-to-end time until EQ-Robin returns the first feasible minimal suite (variant exploration + repeated generation + feasibility checks).
\item \textbf{Variants checked:} number of equivalent variants examined until success.
\item \textbf{Coverage (\%):} achieved MC/DC coverage of the returned suite (100\% indicates all Unique-Cause obligations are satisfied for SBEs).
\item \textbf{Success (1/0):} 1 if a feasible minimal suite is found and achieves 100\% coverage; otherwise 0.
\item \textbf{Speedup (Baseline/EQ):} baseline time divided by EQ time-to-first-feasible.
\item \textbf{EQ overhead (sec):} EQ time-to-first-feasible minus baseline time.
\end{itemize}

\subsection{Experimental Environment}
\label{sec:env}

All experiments were conducted on macOS Tahoe 26.2 with an Apple M4 Pro chip and 24\,GB RAM.
We measure wall-clock time for each run of the pipeline.

\subsection{Results}
\label{sec:results}

Table~\ref{tab:eval_results} reports per-SBE results under the two-condition input-forbidden constraints.
EQ-Robin succeeds on all 25 SBEs (Success $=1$) and achieves 100\% coverage (RQ1--RQ2).
Regarding efficiency (RQ3), EQ-Robin checks on average 2.36 variants (median 2, max 4) before finding the first feasible minimal suite.
Time-to-first-feasible ranges from 0.0718\,s to 1.6611\,s (mean 0.3757\,s; median 0.1772\,s), remaining sub-second in most cases.

\begin{figure*}[t]
\centering
\begin{minipage}[t]{0.49\textwidth}
  \centering
  \includegraphics[width=\textwidth]{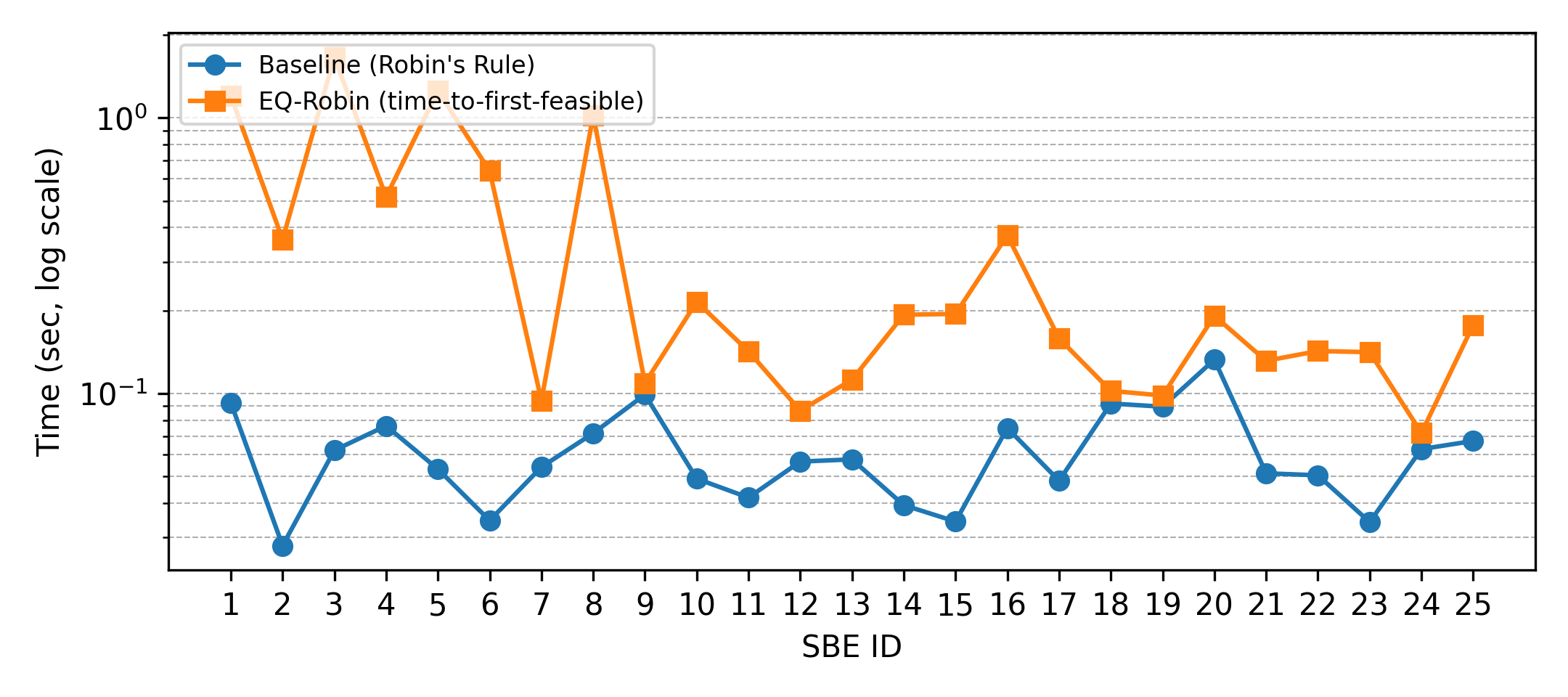}
  \caption{Baseline time vs.\ EQ time-to-first-feasible (log-scale).}
  \label{fig:eval_time}
\end{minipage}\hfill
\begin{minipage}[t]{0.49\textwidth}
  \centering
  \includegraphics[width=\textwidth]{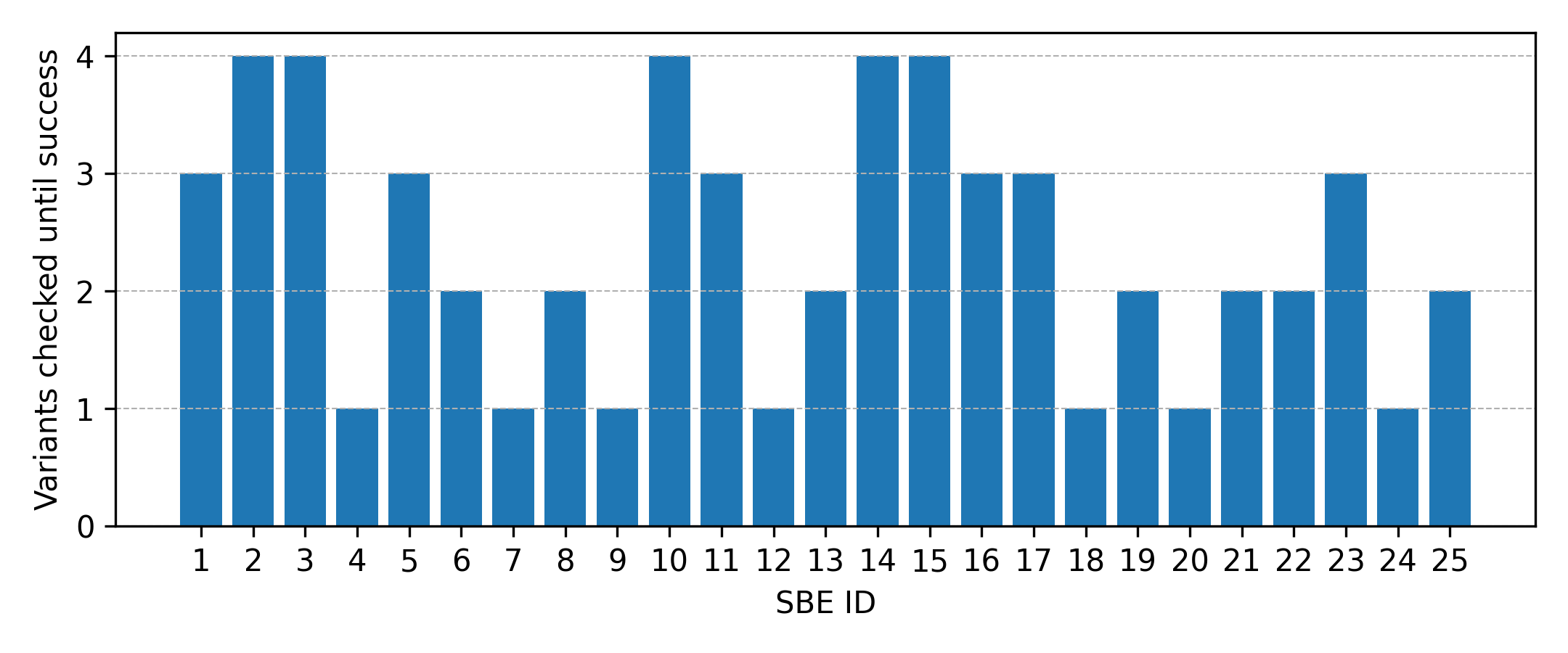}
  \caption{Variants checked until first feasible suite (early-stop).}
  \label{fig:eval_variants}
\end{minipage}
\end{figure*}

Figure~\ref{fig:eval_time} compares baseline generation time against EQ-Robin time-to-first-feasible (log-scale).
As expected, EQ-Robin introduces overhead because it may generate and validate multiple candidate suites until a feasible one is found.
Figure~\ref{fig:eval_variants} visualizes how many variants are required; most SBEs succeed within a small number of checks, which supports the practicality of early-stop exploration.

\begin{table*}[t]
\centering
\scriptsize
\setlength{\tabcolsep}{3.2pt}
\renewcommand{\arraystretch}{1.08}
\caption{Per-SBE results under two-condition input-forbidden constraints (EQ-Robin early-stop mode).}
\label{tab:eval_results}
\resizebox{\textwidth}{!}{%
\begin{tabular}{r r r r r r r r r}
\toprule
ID & $N$ & Baseline time (s) & EQ time-to-first-feasible (s) & Variants checked & Coverage (\%) & Success & Speedup & EQ overhead (s) \\
\midrule
1 & 23 & 0.092557 & 1.202241 & 3 & 100 & 1 & 0.08 & 1.109684 \\
2 & 5 & 0.028010 & 0.361811 & 4 & 100 & 1 & 0.08 & 0.333801 \\
3 & 20 & 0.062134 & 1.661134 & 4 & 100 & 1 & 0.04 & 1.599000 \\
4 & 21 & 0.076227 & 0.516221 & 1 & 100 & 1 & 0.15 & 0.439994 \\
5 & 17 & 0.053131 & 1.253131 & 3 & 100 & 1 & 0.04 & 1.200000 \\
6 & 10 & 0.034554 & 0.641740 & 2 & 100 & 1 & 0.05 & 0.607186 \\
7 & 15 & 0.054029 & 0.094029 & 1 & 100 & 1 & 0.57 & 0.040000 \\
8 & 20 & 0.071639 & 1.021390 & 2 & 100 & 1 & 0.07 & 0.949751 \\
9 & 17 & 0.098991 & 0.108991 & 1 & 100 & 1 & 0.91 & 0.010000 \\
10 & 13 & 0.049013 & 0.215011 & 4 & 100 & 1 & 0.23 & 0.165998 \\
11 & 12 & 0.041849 & 0.142122 & 3 & 100 & 1 & 0.29 & 0.100273 \\
12 & 18 & 0.056544 & 0.086544 & 1 & 100 & 1 & 0.65 & 0.030000 \\
13 & 11 & 0.057589 & 0.112118 & 2 & 100 & 1 & 0.51 & 0.054529 \\
14 & 9 & 0.039290 & 0.192899 & 4 & 100 & 1 & 0.20 & 0.153609 \\
15 & 8 & 0.034258 & 0.194258 & 4 & 100 & 1 & 0.18 & 0.160000 \\
16 & 19 & 0.074857 & 0.374857 & 3 & 100 & 1 & 0.20 & 0.300000 \\
17 & 11 & 0.048177 & 0.158177 & 3 & 100 & 1 & 0.30 & 0.110000 \\
18 & 24 & 0.091853 & 0.102123 & 1 & 100 & 1 & 0.90 & 0.010270 \\
19 & 18 & 0.089465 & 0.098165 & 2 & 100 & 1 & 0.91 & 0.008700 \\
20 & 26 & 0.132569 & 0.191146 & 1 & 100 & 1 & 0.69 & 0.058577 \\
21 & 13 & 0.051261 & 0.131261 & 2 & 100 & 1 & 0.39 & 0.080000 \\
22 & 17 & 0.050424 & 0.142430 & 2 & 100 & 1 & 0.35 & 0.092006 \\
23 & 7 & 0.034108 & 0.141077 & 3 & 100 & 1 & 0.24 & 0.106969 \\
24 & 11 & 0.062772 & 0.071811 & 1 & 100 & 1 & 0.87 & 0.009039 \\
25 & 18 & 0.067169 & 0.177161 & 2 & 100 & 1 & 0.38 & 0.109992 \\
\bottomrule
\end{tabular}}
\end{table*}

\subsection{Threats to Validity}
\label{sec:threats}

\paragraph{Constraint realism}
Our injected constraints use a two-condition forbidden pattern. While simple, it reflects a common industrial situation where particular combinations of condition outcomes are disallowed.
Real systems may impose richer, state-dependent constraints; integrating such oracles (e.g., simulation/hardware checks) is an important direction.

\paragraph{Generality}
We evaluated 25 SBEs; additional studies on larger sets of extracted predicates from real codebases would further strengthen generalizability.


\section{Discussion and Threats to Validity}
\label{sec:discussion}

\subsection{Discussion}
EQ-Robin is designed for an industrial reality where a minimal MC/DC suite can become unusable due to infeasible input combinations.
Our default \emph{early-stop} mode aligns with practice: engineers typically need \emph{one} feasible minimal $(N{+}1)$ suite quickly to proceed with execution and coverage validation in commercial toolchains.
The evaluation indicates that feasibility is often recovered after checking only a few equivalent variants (Table~\ref{tab:eval_results}), suggesting that budgeted equivalence exploration can be practical without exhaustive enumeration.

EQ-Robin is most beneficial when a canonical single-suite construction is fragile under constraints.
By exploring semantically equivalent SBE structures (to which Robin's Rule is structure-sensitive), EQ-Robin can return an alternative minimal suite that preserves coverage while avoiding forbidden patterns.
The current scope focuses on SBEs; extending the approach to non-SBE predicates and richer feasibility interfaces is a natural next step.


\subsection{Threats to Validity}
\label{sec:threats}

\paragraph{Constraint realism}
We evaluate feasibility using a two-condition forbidden-pattern model.
While it captures a common class of industrial restrictions, real systems may impose richer, state-dependent constraints.
Results may differ under more complex feasibility oracles.

\paragraph{Coverage validity}
Coverage is reported based on satisfying Unique-Cause MC/DC obligations for the extracted SBE predicate.
In practice, reported MC/DC from commercial tools may differ if predicate extraction/normalization does not perfectly match the implemented decision logic or if atomic conditions are not independently controllable.

\paragraph{Measurement and randomness}
Wall-clock time can vary across machines and runtime conditions, and the injected constraints depend on random selection of two conditions and values.
We mitigate this by using a fixed environment and fixed seeds, but additional runs across multiple seeds and platforms would strengthen generality.

\paragraph{Generalizability}
We evaluate 25 SBEs from an avionics-style benchmark.
Further studies on predicates extracted from real industrial codebases across domains would improve external validity.


\section{Related Work}
\label{sec:related}

\subsection{MC/DC Variants and Industrial Nuances}
Structural coverage criteria such as MC/DC are widely adopted in safety-critical domains, and the subtle differences among MC/DC variants (e.g., \emph{Unique-Cause} vs.\ \emph{Masking}) have been extensively discussed.\cite{chilenski2001,hayhurst2001}
In practice, the semantics of evaluation (notably short-circuit evaluation in programming languages) can influence how coverage obligations should be interpreted and measured.\cite{kandl2015reasonability}
Beyond semantics, multiple studies report that MC/DC adequacy can be sensitive to implementation structure, such as expression folding/inlining and compilation choices, which may change the effective propagation/observability of faults even when the same logical decision is covered.\cite{rajan2008structure,whalen2008dasc,whalen2013omcdc,gay2015risks}
These observations motivate approaches that do not treat a Boolean expression as a single fixed structure, but instead consider how \emph{equivalent structures} and instrumentation/oracle choices affect practical coverage and fault-revealing ability.

\subsection{Automated MC/DC-Oriented Test Generation}
Automated test generation from Boolean specifications predates modern solver-based pipelines; early work investigated how to derive test data directly from Boolean specifications without assuming any single, fixed program structure.\cite{weyuker1994}
A major research line encodes coverage obligations (including MC/DC-like criteria) into formal constraints and uses automated reasoning engines to construct test sequences or test vectors.\cite{rayadurgam2001coverage,rayadurgam2003sew}
In this direction, bounded model checking and verification-based techniques have also been used to generate tests or to compute strict coverage scores with strong correctness arguments.\cite{golla2024,golla2025}

Complementary to model checking, solver-based approaches encode MC/DC obligations as SAT/SMT constraints and search for minimal or near-minimal suites.\cite{kitamura2018,yang2018,jaffar2019}
These approaches are powerful and naturally accommodate additional domain constraints (e.g., feasibility rules expressed in OCL-like constraint languages).\cite{sartaj2019ocl,sartaj2025}
However, when constraints eliminate some required test vectors, these methods may return non-minimal suites, incomplete coverage, or require substantial solver guidance/tuning.

Graph-based formulations provide another path: decision diagrams (e.g., BDDs/ROBDDs) can compactly represent Boolean decisions and have been used to extract MC/DC-adequate (and sometimes minimal) suites by exploring path structure and independence pairs.\cite{ahishakiye2021bdds}
Yet, even with BDD-based selection, strongly coupled conditions or domain restrictions can prevent realizing the theoretical minimal size (often discussed as $(N{+}1)$ for $N$ conditions) in a \emph{single} feasible suite.\cite{chilenski2001,hayhurst2001,ahishakiye2021bdds}

Finally, heuristic and search-based strategies—including coverage-guided meta-heuristics and learning-based generation—aim at scalability and multi-objective optimization (e.g., effort, runtime, fault detection proxies).\cite{barisal2021,kaur2011,cegin2020,cao2024}
Tool-oriented efforts additionally propose end-to-end MC/DC test-data pipelines for engineering usage.\cite{shekhawat2021,haque2014,ghani2009,awedikian2009}
While effective in many settings, such methods typically do not provide a deterministic guarantee of producing a minimal $(N{+}1)$ \emph{Unique-Cause} suite for SBEs under arbitrary industrial feasibility constraints.

\subsection{Positioning of EQ-Robin}
Our prior work (Robin's Rule) deterministically constructs a minimal-size $(N{+}1)$ test suite for SBEs under the assumption that required test vectors are feasible.\cite{lee2025robin}
The industry gap arises when some vectors are infeasible due to input restrictions, environment constraints, sensor/actuator limitations, or setup costs—causing coverage to drop even though the constructed suite is theoretically minimal.
EQ-Robin addresses this gap by exploring a \emph{design space of semantically equivalent expression structures} and generating a \emph{family} of minimal suites, enabling engineers to select a feasible suite under industrial constraints.
In contrast to purely stochastic search, EQ-Robin preserves a deterministic minimal-suite constructor as the core engine, and uses equivalence exploration plus secondary ranking (cost models) to support practical selection without sacrificing interpretability.


\section{Conclusion}
\label{sec:conclusion}

This paper presented EQ-Robin, a constraint-resilient extension of Robin's Rule for Singular Boolean Expressions.
While Robin's Rule deterministically produces a minimal $(N{+}1)$ Unique-Cause MC/DC suite, industrial feasibility constraints can invalidate the canonical suite.
EQ-Robin addresses this gap by exploring semantically equivalent SBE structures and returning the \emph{first feasible} minimal suite via early stop.
Our evaluation on 25 SBEs under input-forbidden constraints shows that EQ-Robin consistently recovers feasible minimal suites with 100\% coverage, typically after checking only a small number of variants.
Future work includes integrating richer feasibility oracles and extending the approach beyond SBEs.


\appendix
\section{Benchmark SBEs (IDs 1--25)}
\label{app:sbes}

\small
\sloppy
This appendix lists the canonical SBE predicates used in the evaluation (IDs 1--25).
We present each predicate as a C/Boolean-style expression. Long expressions may wrap across lines.

\begin{enumerate}
\item[\textbf{ID 1.}]  \texttt{(!(a \&\& b) \&\& (c \&\& !d \&\& !e || !f \&\& g \&\& !h || !i \&\& !j \&\& !k) \&\& (l \&\& m \&\& (n || o) \&\& p || q \&\& (r || s) \&\& !t || u \&\& (v || w)))}
\item[\textbf{ID 2.}]  \texttt{(a \&\& (!b || !c) \&\& d || e)}
\item[\textbf{ID 3.}]  \texttt{a \&\& (!b || !c || d \&\& e \&\& !(!f \&\& g \&\& h \&\& !i || !j \&\& k \&\& l) \&\& !(!m \&\& n \&\& o \&\& p || !q \&\& !r \&\& s)) || t}
\item[\textbf{ID 4.}]  \texttt{(!a \&\& b || c \&\& !d) \&\& !(e \&\& f) \&\& !(g \&\& h) \&\& !(i \&\& j) \&\& (( k \&\& l || m \&\& n ) \&\& o \&\& (!p || !q \&\& !r || !s \&\& (!t || !u)))}
\item[\textbf{ID 5.}]  \texttt{(!a \&\& b || c \&\& !d) \&\& !(e \&\& f) \&\& !(g \&\& h) \&\& ((i \&\& j || k \&\& l) \&\& m \&\& (n \&\& o || !p \&\& q))}
\item[\textbf{ID 6.}]  \texttt{!(a \&\& b) \&\& (!c \&\& d \&\& !e \&\& !f \&\& (g \&\& h || !i \&\& j))}
\item[\textbf{ID 7.}]  \texttt{a \&\& !b \&\& !c \&\& d \&\& !e \&\& f \&\& (g || !h \&\& (i || j)) \&\& !(k \&\& l || !m \&\& n || o)}
\item[\textbf{ID 8.}]  \texttt{a \&\& !b \&\& !c \&\& !((d \&\& (e || !f \&\& (g || h))) || i \&\& (j || !k \&\& (l || m)) \&\& !n \&\& !o ) \&\& !(p \&\& q || !r \&\& s \&\& !t)}
\item[\textbf{ID 9.}]  \texttt{a \&\& !b \&\& !c \&\& (d \&\& (e || !f \&\& (g || h)) \&\& (!i \&\& !j || k) || !l) \&\& (m \&\& n || !o \&\& p \&\& !q)}
\item[\textbf{ID 10.}] \texttt{a || b || c || !d \&\& !e \&\& f \&\& g \&\& !h \&\& !i || j \&\& (k || l) \&\& !m}
\item[\textbf{ID 11.}] \texttt{a \&\& b \&\& (c || d) \&\& e || f \&\& (g || h) \&\& !i || j \&\& (k || l)}
\item[\textbf{ID 12.}] \texttt{a \&\& ((b || c || d) \&\& e || f \&\& g || h \&\& (i || j || k || l)) || (m || n) \&\& (o || p || q) \&\& r}
\item[\textbf{ID 13.}] \texttt{(a \&\& b || c \&\& d) \&\& e \&\& (f || (g \&\& (h \&\& i || j \&\& k)))}
\item[\textbf{ID 14.}] \texttt{(a \&\& b || c \&\& d) \&\& e \&\& (f \&\& g || !h \&\& i)}
\item[\textbf{ID 15.}] \texttt{!a \&\& b \&\& !c \&\& !d \&\& (e \&\& f || !g \&\& h)}
\item[\textbf{ID 16.}] \texttt{(((a || b) || c \&\& !(!(d \&\& e)) \&\& f || g) || !h || i || !((j || !(!((k \&\& (l \&\& !m))) || n \&\& o || p || q \&\& r || s))))}
\item[\textbf{ID 17.}] \texttt{((a \&\& b \&\& c \&\& d) \&\& !(e || f) || (!(!g) \&\& !h \&\& i) \&\& j \&\& k)}
\item[\textbf{ID 18.}] \texttt{!(a || b) || (c \&\& d \&\& !e || !f \&\& g || (!h || i) \&\& j \&\& !k) || (l \&\& m \&\& (n || o) \&\& p || q \&\& (r || s) \&\& !t \&\& u \&\& (v \&\& w) || x)}
\item[\textbf{ID 19.}] \texttt{!(a || (b \&\& !c)) \&\& (!(!(d \&\& e)) \&\& f || g || (!((h || i) || !j) \&\& !(!k))) || !(!(l \&\& m)) \&\& n \&\& o \&\& !p || (!q || r)}
\item[\textbf{ID 20.}] \texttt{((!((a \&\& b)) \&\& !((c \&\& (!(!d) || e \&\& f)) \&\& ((!g \&\& (!h || !(!i) || (j || k))) \&\& !l || m) \&\& n || !(!o) || p)) || !(!((q || r || s) || !t \&\& !((!(!u) || v || w)) || !(!x) \&\& !y \&\& z)))}
\item[\textbf{ID 21.}] \texttt{(a || !b || !(c \&\& d) || (e \&\& f) || g) \&\& (h \&\& i \&\& j) || (k \&\& l \&\& m)}
\item[\textbf{ID 22.}] \texttt{a || b \&\& c || d \&\& e || (!(!(f || g \&\& h) || i \&\& !(!j) \&\& !k) \&\& l || m || !n \&\& o \&\& p \&\& q)}
\item[\textbf{ID 23.}] \texttt{(!a \&\& b || c) || d \&\& e || !f \&\& g}
\item[\textbf{ID 24.}] \texttt{a \&\& !b \&\& c || d \&\& !e \&\& !((f \&\& g || (!h \&\& i)) \&\& j || k)}
\item[\textbf{ID 25.}] \texttt{!((a \&\& b)) \&\& c || (!(((d || e) || f) \&\& g) \&\& h \&\& !(!(i \&\& j))) || k \&\& l || m || n || o \&\& (p \&\& q \&\& r)}
\end{enumerate}

\normalsize


\end{document}